\documentstyle[twoside,fleqn,espcrc2]{article}
\newcommand{\be}{\begin{eqnarray}}
\newcommand{\ee}{\end{eqnarray}}
\newcommand{\dbar}{\mbox{{\rm d\hspace{-2truemm}$^-$}}}

\newcommand{\expec}[1]{\mbox{$\langle\, #1\,\rangle$}}

\newcommand{\AmS}{{\protect\the\textfont2
  A\kern-.1667em\lower.5ex\hbox{M}\kern-.125emS}}
\title{Quantized shells as a tool for studying semiclassical
effects in general relativity}
\author{Roberto Casadio
\address{Dipartimento di Fisica, Universit\`a di Bologna and\\
Istituto Nazionale di Fisica Nucleare, Sezione di Bologna\\
via Irnerio 46, 40126 Bologna, Italy \\
e-mail: casadio@bo.infn.it}}
\begin{document}
\begin{abstract}
Thin shells in general relativity can be used both as models of
collapsing objects and as probes in the space-time outside compact
sources.
Therefore they provide a useful tool for the analysis of the final
fate of collapsing matter and of the effects induced in the matter
by strong gravitational fields.
We describe the radiating shell as a (second quantized) many-body
system with one collective degree of freedom, the (average) radius,
by means of an effective action which also entails a thermodynamic
description.
Then we study some of the quantum effects that occur in the matter
when the shell evolves from an (essentially classical) large initial
radius towards the singularity and compute the corresponding
backreaction on its trajectory.
\end{abstract}
\maketitle
\section{Introduction}
A compact object which collapses gravitationally can be used
as a probe to test physics at different length scales.
In fact, as long as the collapse proceeds, the characteristic size of
the object will subsequently cross all the relevant scales, which
include the Compton wavelengths $\ell_\phi$ of the particle excitations
the object is made of and, eventually, the Planck length $\ell_p$.
Starting with a large size, the dynamics can be initially described by
classical general relativity, therefore avoiding the issue of the
initial conditions which plagues the case of the expansion
({\em e.g.}, of the universe in quantum cosmology).
Then, the quantum nature of the matter in the object will
become relevant presumably around $\ell_\phi$ and a quantum theory
of gravity is required near $\ell_p$.
\par
In this picture one must also accommodate for the role played by
the causal horizon generated by the collapsing object.
The horizon is a sphere with radius $R_G=2\,M$ determined by the total
energy $M/G$ of the system \cite{mtw} ($G$ is the Newton constant).
If the object does not lose energy, $M$ is constant in time and one
may simply take the point of view of a (proper) observer comoving with
some particle of the infalling matter, for whom the horizon is totally
harmless.
The situation is rather different for a radiating object, since one can
then consider also an external observer who witnesses the collapse
by detecting the emitted radiation\footnote{We conceive an observer
as some material device and not as an abstract concept.}.
For such an observer $R_G$ is the radius of a (apparent) horizon which
shrinks in time with an uncertainty in location related to the
probabilistic (quantum mechanical) nature of the emission process
(see \cite{cv} for an analogy with the Unruh effect).
Further, there is an uncertainty (of order $\ell_\phi$) in the position
of the particles of the collapsing body and this all renders the
issue of (viewing) an emitting source crossing its own horizon of
particular relevance, both for the theoretical understanding
of gravitation and for astrophysical applications.
\par
It seems appropriate to tackle this hard topic by considering simple
cases which capture the basic features of the problem and reduce the
technical difficulties.
Such an example is given by the spherically symmetric distributions of
matter with infinitesimal thickness known in general relativity as thin
shells \cite{mtw,israel}.
They can either be pulled just by their own weight or fall in the
external gravitational field of a (spherical) source placed at their
centre, the latter case being of interest for studying the accretion
of matter onto a black hole \cite{mtw} and the Hawking effect
\cite{hawking}.
\section{Effective action}
The equations of motion for thin shells are usually obtained in general
relativity from the junction conditions between the embedding four-metric
and the intrinsic three-metric on the surface of the shell of radius $r=R$
\cite{israel,mtw}.
Their dynamics can also be described by an effective action \cite{acv}
which is obtained by inserting in the general expression for the
Einstein-Hilbert action the Schwarzschild solution \cite{mtw} with
mass parameter $M_0$ ($\ge 0$) as the space-time inside the shell ($r<R$)
and the Vaidya solution \cite{vaidya} with mass function $m$ ($>M_0$)
outside the shell ($r>R$).
When $\dot M$, the time derivative of $m$ evaluated on the outer surface
of the shell, is negative, the shell emits null dust.
\par
The effective action is given by \cite{acv}
\be
S_s&=&\int{dt\over G}\,
\left[R\,\beta-R\,\dot R\,\tanh^{-1}\left({\dot R\over\beta}\right)
\right]_{out}^{in}
\nonumber \\
&&-\int dt\,N\,E
\nonumber \\
&&+\int{dt\over 2\,G}\,\left[{\dot M\,R^2\over R-2\,M}
-{M\,R\,\dot R\over R-2\,M}\right]
\ ,
\label{Seff1}
\ee
and is a functional of $R$, $M$ and $N$ (lapse function of the three-metric
on the shell); $E=E(R,M)$ is the shell energy,
$\beta^2=(1-2\,m/R)\,N^2+\dot R^2$ and $[F]^{in}_{out}=F_{in}-F_{out}$
denotes the jump of the function $F$ across the shell.
The first two integrals in (\ref{Seff1}) were known from an analogous
derivation for non-radiating shells \cite{guth} and the third integral
properly accounts for the fact that a radiating shell defines an open
(thermo)dynamical
system\footnote{This term becomes dynamically irrelevant when
the radiation ceases \cite{acv}.}.
\subsection{Equations of motion}
The Euler-Lagrange equations of motion,
\be
{\delta S_s\over\delta N}&\equiv&-[H_G+E]=0
\label{H}
\\
{\delta S_s\over\delta R}&=&0
\label{P}
\\
{\delta S_s\over\delta M}&\equiv&K=0
\label{dM}
\ ,
\ee
represent, respectively, the (primary) Hamiltonian constraint
corresponding to
the time-reparametrization invariance of the shell, an equation
for the surface tension $P=\partial E/\partial A$
($A=4\,\pi\,R^2$ is the area of the surface of the shell) and a
second (primary) constraint which relates $\dot M$ to
the luminosity of the shell.
The equations (\ref{H}) and (\ref{P}) coincide with the standard
junction conditions \cite{israel}, but (\ref{dM}) only appears in
this approach \cite{acv}.
\par
Since from (\ref{H})-(\ref{dM}) it follows that
\be
\dot H_G+\dot E&\sim& -\dot A\,{\delta S_s\over\delta A}
-\dot M\,{\delta S_s\over\delta M}=0
\label{dH}
\\
\dot K\sim\ddot H&=&0
\ ,
\ee
no new (secondary) constraint arises and the dynamical system
defined by $S_s$ is consistent.
One can therefore set $N=N(t)$ and $M=M(t)$ any fixed function
of the time and correspondingly solve for $R=R(t)$.
\subsection{Thermodynamics}
The meaning of the identity (\ref{dH}) is that the total energy of the
system is conserved and it can also be cast in the form of the first
principle of thermodynamics \cite{acv},
\be
dE=P\,dA+\dbar Q
\ ,
\ee
where $\dot Q\sim\dot M$ is the luminosity.
A second principle can also be introduced, at least in the quasi-static
limit $\dot R^2\ll 1-2\,M/R$, by defining an entropy $S$ and a temperature
$T$ such that
\be
dS={\dbar Q\over T}
\ee
is an exact differential, which yields \cite{acv}
\be
T&=&{a\over 8\,\pi\,k_B}\,{\ell_p^{1-b}\over
\left(\hbar\,M\right)^{b}}
\,{1\over\sqrt{1-2\,M/R}}
\nonumber \\
&\equiv& {T_{a,b}\over\sqrt{1-2\,M/R}}
\ ,
\ee
where $k_B$ is the Boltzmann constant, $a$ and $b$ are constants
and $T/T_{a,b}$ is the Tolman factor.
We note in passing that $T_{1,1}$ is the Hawking temperature of a black
hole of mass $M$ \cite{hawking}.
\subsection{Microstructure}
A microscopic description of the shell can be obtained by
considering $n$ close microshells of Compton wavelength
$\ell_\phi=\hbar/m_\phi$ \cite{acvv}.
One then finds that such a (many-body) system is gravitationally
confined within a thickness $\Delta$ around the mean radius $R$
\footnote{This is the reason we refer to $R$ as a collective
variable.}
and can estimate $\Delta$ from an Hartree-Fock approximation for
the wavefunction of each microshell.
This yields \cite{acvv}
\be
\left({\Delta\over R}\right)^{3/2}
\sim G\,n\,\hbar\,m_\phi\,{\ell_\phi\over R^2}
\sim{\ell_\phi\,R_G\over R^2}
\ .
\label{delta}
\ee
which, for $R\ge R_G$, is negligibly small provided
\be
R\gg\ell_\phi
\ ,
\label{check}
\ee
in agreement with the naive argument that the location of an object
cannot be quantum mechanically defined with an accuracy higher than
its Compton wavelength.
\par
In the limit (\ref{check}) one can second quantize the shell by
introducing a (scalar) field $\phi$ with support within a width
$\Delta$ around $R$ and Compton wavelength $\ell_\phi$ \cite{acvv}
and obtains (neglecting terms of order $\Delta$ and higher)
\be
E&\simeq&
{1\over2\,\ell_\phi}\,\left[{\pi_\phi^2\over R^2}+R^2\,\phi^2\right]
\nonumber \\
&&+H_{int}(\phi,M,\dot M,R)
\ ,
\ee
where $\pi_\phi$ is the momentum conjugated to $\phi$ and $H_{int}$
describes the local interaction between the matter in the shell and
the emitted radiation.
When $H_{int}\not=0$, one expects that $M=M(t)$ becomes a dynamical
variable which cannot be freely fixed and the luminosity should then
be determined by the corresponding Euler-Lagrange equation (\ref{dM})
from purely initial conditions for $R$ and $M$ (in any gauge $N=N(t)$)
\cite{acvv2}.
\section{Semiclassical description}
Lifting the time-reparametrization invariance of the shell to a
quantum symmetry yields the Wheeler-DeWitt equation \cite{dewitt}
corresponding to the classical Hamiltonian constraint (\ref{H}).
For the non-radiating case ($\dot M=H_{int}=0$) and in the proper
time gauge ($N=1$) it is given by \cite{acvv}
\be
\left[\hat H_G(P_R,R)+\hat E(P_\phi,\phi;R)\right]\Psi=0
\ .
\label{wdw}
\ee
\par
One can study (\ref{wdw}) in the Born-Oppeheimer approach
\cite{bfv} by writing
\be
\Psi(R,\phi)&=&\psi(R)\,\chi(\phi;R)
\nonumber \\
&\simeq&\psi_{WKB}(R)\,\chi(\phi;R)
\ ,
\label{wkb}
\ee
where $\psi_{WKB}$ is the semiclassical (WKB) wavefunction for the
radius of the shell.
This allows one to retrieve the semiclassical limit in which
$R$ is a collective (semi)classical variable driven by the
expectation value of the scalar field Hamiltonian operator
$\hat E$ over the quantum state $\chi$,
\be
H_G+\expec{\hat E}=0
\ ,
\label{R}
\ee
while $\chi$ evolves in time according to the Schr\"odinger equation
\be
i\,\hbar\,{\partial\chi\over\partial t}=\hat E\,\chi
\ .
\label{schro}
\ee
\par
In general, in order to obtain (\ref{R}) and (\ref{schro}) from
(\ref{wdw}), one needs to assume that certain fluctuations
(corresponding to quantum transitions between different trajectories
of the collective variable $R$) be negligible \cite{bfv}.
For the present case one can check {\em a posteriori} that this is
true if the condition (\ref{check}) holds (that is, $n$ is sufficiently
big, see (\ref{delta})) \cite{acvv}.
\subsection{Particle production and backreaction}
The equation (\ref{schro}) can be solved beyond the adiabatic
approximation by making use of {\em invariant} operators \cite{lewis}.
In particular, one can choose $\chi$ as the state with initial (proper)
energy $E_0=m_0/G$ and radius $R_0$ and expand in the parameter of
non-adiabaticity $\delta^2=\ell_\phi^2\,M/R_0^3$ to obtain
(to first order in $\delta^2$) \cite{acvv}
\be
m=G\,\expec{\hat E}\simeq
m_0\,\left(1+{R_0^3\,\dot R^2\over R_G\,R^2}\,\delta^2\right)
\ ,
\label{E}
\ee
which is an increasing function for decreasing $R$.
This signals a (non-adiabatic) production of matter particles in the
shell along the collapse.
Since the total energy $M/G$ is conserved, such a production can be
viewed as a transfer of energy from the collective degree of freedom
$R$ to the microscopic degree of freedom $\phi$ and one therefore
expects a slower approach towards the horizon.
\par
In fact, the equation of motion (\ref{R}) for $R$,
\be
\dot R^2={R_G\over 2\,R}+2\,\left(1-{2\,m\over R_G}\right)
-{\ell_\phi^2\,m^2\over R_G^2\,R^2}
\ .
\label{R_s}
\ee
can be integrated numerically along with (\ref{E}) to compute
$m=m(t)$ and the corresponding backreaction on the trajectory $R=R(t)$
confirms the above qualitative argument \cite{acvv}.
\subsection{Gravitational fluctuations}
One can also study the effects due to higher WKB order terms in the
gravitational wavefunction by defining
\be
\psi(R)=f(R)\,\psi_{WKB}(R)
\ ,
\ee
where $f$ must then satisfy \cite{ebo}
\be
i\,\hbar\,{\partial f\over\partial t}\simeq
\hbar\,{\ell_p^2\over 2\,R}\,
\left.{\partial^2 f\over \partial R^2}\right|_{R_c}
\ ,
\label{sc0}
\ee
in which $R_c=R_c(t)$ is the (semi)classical trajectory $\psi_{WKB}$ is
peaked on and we have neglected terms of order $\dot R/R$ and
higher.
\par
Acceptable solutions \cite{ebo} to (\ref{sc0}) are given by plane
waves with wave numbers $\lambda\ge\ell_p$ and negative ``energy''
\be
E_\lambda=-{\hbar\over 2\,R_c}\,{\ell_p^2\over\lambda^2}
\ ,
\label{Ela}
\ee
which agrees with the fact that the gravitational contribution to the
total (super)Hamiltonian has the opposite sign with respect to matter.
\par
Another basic feature of (\ref{Ela}) is that $E_\lambda$ is proportional
to $R_c^{-1}$, which makes it negligible with respect to $\expec{\hat E}$
for large radius, but one then expects appreciable corrections as
the collapse proceeds.
One can indeed take $E_\lambda$ into account in an improved
semiclassical equation \cite{ebo} for the trajectory of the shell
\be
H_G+\expec{\hat E}+E_\lambda=0
\ ,
\ee
which predicts a breakdown of the semiclassical approximation for values
of $R$ larger than the limit (\ref{check}) and possibly larger than
$R_G$.
This would imply that the whole shell becomes a quantum object
far before reaching the (quantum mechanically unacceptable)
singularity $R=0$.
\end{document}